\documentclass[12pt]{article}
\usepackage[margin=1in]{geometry}
\usepackage{setspace}
\usepackage{mathtools}
\usepackage{amssymb}
\usepackage{braket}
\usepackage{amsmath,amsthm}
\usepackage{authblk}
\usepackage[english]{babel}
\usepackage{xcolor}
\usepackage{mathrsfs}

\usepackage{verbatim}
\usepackage{enumitem}
\begin{document}
\begin{center}	
	\begin{Large}
		\textbf{Conservation in Every Quantum Measurement Outcome?}\\
	\end{Large}
\end{center}
	
\begin{center}
	\begin{large}
		E. Okon, D. Sudarsky and M. Wiedemann\\
	\end{large}
	\textit{Universidad Nacional Aut\'onoma de M\'exico, Mexico City, Mexico.}\\[1cm]
\end{center}
In quantum mechanics, conservation laws are usually understood only statistically and only in the context of unitary evolution: during such evolution, if an observable commutes with the Hamiltonian, its expectation value and Born-rule probability distribution remain unchanged. Individual measurement outcomes, however, disrupt even this weak form of statistical conservation. Recently, though, it has been argued that a stronger, non-statistical form of conservation, valid at the level of individual measurement outcomes, can be recovered. The proposal relies on the claim that any superposition of eigenstates of a conserved quantity must ultimately be prepared relative to another physical system, which serves as the relevant reference frame and remains correlated with the system being prepared. On this enlarged description, the change observed in the measured system is claimed to be exactly compensated by an opposite change in the preparer, thereby restoring conservation at the level of individual measurement outcomes. We analyze this proposal and argue that it faces significant conceptual difficulties. We show that the proposed mechanism is not adequately justified and, moreover, does not establish conservation in individual measurement outcomes as claimed.

\onehalfspacing
\section{Introduction} 

In quantum mechanics, an observable represented by a Hermitian operator $\hat O$ is usually said to be a constant of motion whenever it commutes with the Hamiltonian, $[\hat O,\hat H]=0$. Under unitary evolution, this guarantees that the expectation value of $\hat O$, as well as the Born probabilities associated with its possible values, remain unchanged in time. This, however, is already a substantially weaker notion than that expressed by conservation laws in classical physics. In the classical case, a conservation law normally concerns a quantity possessed by an individual physical system whose value remains fixed throughout the relevant evolution. By contrast, the quantum condition $[\hat O,\hat H]=0$ need not imply that the system possesses any determinate value of $O$ at all, much less one that persists through time (see \cite{maudlin2020status}). Nevertheless, preservation of expectation values or, more strongly, of the probability distribution associated with $\hat O$ is often presented as the quantum-mechanical counterpart of a conservation law.

Even this weaker notion encounters an apparent difficulty once measurements are taken into account. A system may initially be in a superposition of different eigenstates of a conserved observable and, upon measurement, be found with one definite value. Conditioned on the particular outcome obtained, both the state and the corresponding probability distribution have changed. Thus, while unitary dynamics preserves the distribution of a conserved observable, the usual measurement postulates fail to preserve conservation in each individual realization.

In \cite{collins2025conservation}, however, it is argued that this conclusion results from describing the measurement process too narrowly. According to them, quantum conservation laws can be understood as holding not merely statistically but in every individual measurement outcome. Their central claim is that the apparent violation arises because the usual description neglects the physical system involved in \emph{preparing} the initial superposition. The preparation process leaves the measured system correlated, even if only very weakly, with this ``preparer.'' When a measurement subsequently yields a definite value of the conserved quantity, the corresponding change in the system is, they argue, exactly compensated by an opposite change in the preparer. On this enlarged description, the total conserved quantity of the system plus preparer is claimed to remain unchanged in each individual outcome. Moreover, the authors emphasize that the bookkeeping can be localized to this restricted collection of systems; conservation need not be restored by appealing to unspecified degrees of freedom elsewhere in the universe, nor does the argument require the universe as a whole to possess a definite value of the conserved quantity.

The proposal is attractive because it aims to recover a notion of conservation stronger than the usual ensemble-level one while remaining entirely within quantum mechanics. We contend, however, that it faces difficulties at two distinct levels. First, we question the justification offered for the framework itself. In particular, we argue that the appeal to physical reference frames does not establish that every superposition must be accompanied by an additional material system, that the specific preparation interaction required by the mechanism is postulated rather than shown to be necessary, and that recursively applying the same reasoning to preparers generates further conceptual difficulties. We also argue that the proposed extension to several noncommuting conserved quantities is problematic. Second, independently of these justificatory concerns, we examine whether the mechanism actually establishes the stronger form of conservation claimed. We argue that the formalism appears to preserve, at most, a suitably localized version of the usual statistical conservation law and that, in the case of energy, the interaction required for preparation introduces an additional contribution that prevents the proposed subsystem bookkeeping from straightforwardly establishing conservation.

The paper is organized as follows. In Section~\ref{prop}, we reconstruct the proposal of \cite{collins2025conservation}, emphasizing the role of the preparer, the reference-frame argument, the preparation interaction, the chain of preparers, and the proposed extension to noncommuting conserved quantities. In Section~\ref{As}, we assess these elements in turn. We first examine the justification of the framework and then consider whether, even granting its central assumptions, the mechanism succeeds in establishing conservation at the level of individual measurement outcomes. Section~\ref{Con} summarizes our conclusions.

\section{A proposal for conservation in every quantum measurement}
\label{prop}

In \cite{collins2025conservation}, an argument intended to restore conservation laws at the level of every individual measurement outcome is proposed. Although their analysis is developed explicitly for angular momentum in two dimensions, the authors argue that the underlying logic should extend to other conserved quantities, including energy. Indeed, their work is partly motivated by earlier discussions concerning energy conservation in quantum mechanics 
\cite{aharonov2021conservation,
aharonov2023conservation}. In what follows, we outline the core elements of their proposal.

Consider a system $S$ in a superposition of eigenstates of a conserved observable
\begin{equation}
\label{Sup}
\ket{\psi}_S=\sum_n \alpha(n) \ket{n}_S.
\end{equation}
As mentioned above, while the probability distribution of this observable is conserved at the ensemble level, the standard formulation does not provide conservation at the level of individual measurement outcomes. In \cite{collins2025conservation}, it is argued, however, that this description is fundamentally incomplete. They first consider a particular physical preparation procedure in which an external system \(P\), the ‘preparer,’ generates the desired superposition while conserving the total quantity. Specifically, the system and preparer start in an initially uncorrelated state:
\begin{equation}
\ket{\Psi_0}=|\Phi\rangle_P \otimes |0\rangle_S
\end{equation} 
where $\ket{\Phi}_P$ denotes the initial state of the preparer and $\ket{0}_S$ is a reference eigenstate for the relevant conserved quantity of the system. The preparation is then implemented through a particular conservation-preserving unitary transformation,
\begin{equation}
\label{UT}
\ket{\Psi_0}\longrightarrow \ket{\Psi'}=\sum_n \alpha(n) |\Phi-n\rangle_P \otimes|n\rangle_S.
\end{equation}
Here \(\ket{\Phi-n}_P\) denotes the state of the preparer shifted by precisely the amount of the conserved quantity acquired by the system. The subsequent measurement is modeled by an interaction that correlates the system with a pointer while leaving the measured conserved quantity unchanged. In the simple model considered, the measuring device therefore does not supply the compensating amount; the relevant compensation is already encoded in the correlations established with the preparer.

That is, every component $\ket{n}_S$ is correlated with a compensating state of the preparer carrying the opposite change. 
The crucial observation is that conservation now holds for the composite system $S+P$. If a subsequent measurement of the conserved observable yields the outcome $k$, the post-measurement state becomes
\begin{equation}
\ket{\Psi'}\longrightarrow \ket{\Psi^{(k)}}=\ket{\Phi-k}_P\otimes\ket{k}_S.
\end{equation}
Then, for each outcome \(k\), the shift \(+k\) in the system is accompanied by a corresponding shift \(-k\) in the preparer, so that the total value associated with that branch matches the corresponding initial total.

Thus, what appears from the perspective of the system alone as a random jump between different values is, from the perspective of the enlarged system $S+P$, merely the selection of one component of a previously existing entangled structure that upholds the conservation law in every component. 

To recover the initial description of \(S\) as approximately pure, the shifted preparer states \(|\Phi-n\rangle_P\) must have nearly unit mutual overlap. This can be achieved by choosing the preparer to have a sufficiently broad distribution in the conserved quantity (which allows for a narrow distribution in its conjugate variable). As a result, after tracing out the preparer, the reduced density matrix of the system can be made arbitrarily close to the pure state, while still retaining a small but crucial residual entanglement with the preparer that guarantees conservation in every individual measurement outcome.

The upshot is that, in the enlarged description, the change in the conserved quantity of the measured system is exactly compensated by an opposite change in the corresponding quantity of the preparer, thereby restoring conservation at the level of individual measurement outcomes. Moreover, this conservation is claimed to be localizable, in the sense that it can be accounted for by considering only the system and the preparer, without any appeal to the rest of the universe.

This framework raises three issues that the authors explicitly address, together with a fourth issue concerning its generalization beyond the simplified model considered:
\begin{itemize}
\item[(i)] What happens if a system is simply ``given'' in a pure, isolated superposition of eigenstates of a conserved quantity, with no residual entanglement with any preparer?
\item[(ii)] What guarantees that the specific preparation interaction given in Eq.~(\ref{UT}) must actually occur?
\item[(iii)] If the preparer is itself in a coherent superposition, it must in turn have been prepared by a preceding system. Does the compensating transfer of the conserved quantity then propagate recursively through an infinite chain of preparers?
\item[(iv)] Can the proposal be extended to several noncommuting conserved quantities?
\end{itemize}

The proposed resolutions of (i) and (ii) both rely on the role of reference frames. We examine the authors' treatment of these two issues in the following two subsections, before turning to points (iii) and (iv) in subsections \ref{GP} and \ref{SN}, respectively.

\subsection{Preparers}

To address (i), in \cite{collins2025conservation} it is argued that a quantum system can never be ``given" in an isolated superposition of a conserved quantity. For example, in the case of angular momentum, a coherent superposition, such as the one in Eq.~(\ref{Sup}), implicitly defines a non-uniform angular distribution on the circle. But, the authors argue, the same angular distribution could be placed in a different location, rotated around the circle from the first, by appropriately changing the phases. Thus the phase is related to the angle. However, an absolute angle is meaningless: angle only has meaning relative to a frame of reference $F$, \cite{aharonov1967charge,aharonov1984quantum,bartlett2007reference}. 
Consequently, a coherent superposition of angular momentum must be prepared relative to the frame's angular position $\theta_f$, leading to a joint state of the kind
\begin{equation}
\label{JS}
\int \tilde{\Phi}_f(\theta_f)|\theta_f\rangle_F \int \tilde{\alpha}(\theta_s - \theta_f) |\theta_s\rangle_S d\theta_sd\theta_f
\end{equation}
The function \(\tilde{\alpha}(\theta_s-\theta_f)\) is the Fourier transform of the coefficients \(\alpha(n)\), so Eq.~(\ref{JS}) is simply the angle-space representation of the superposition \(\sum_n \alpha(n) \ket{n}_S\), written relative to the frame angle \(\theta_f\).

\subsection{Preparation interaction}

To address (ii), the authors then argue that this relational state is the exact mathematical outcome of the specific preparation interaction introduced in Eq.~(\ref{UT}) when expressed in the conjugate continuous angular representation
\begin{align}
\int \tilde{\Phi}_p(\theta_p)|\theta_p\rangle_P d\theta_p\, |0\rangle_S \longrightarrow
\int \tilde{\Phi}_p(\theta_p)|\theta_p\rangle_P \int \tilde{\alpha}(\theta_s-\theta_p) |\theta_s\rangle_S d\theta_sd\theta_p
\end{align}
The authors therefore identify the preparer with the physical reference frame relative to which the superposition is defined. They do not claim that the immediate preparation of \(S\) must always be given by Eq.~(\ref{UT}). Rather, they postulate that somewhere in the preparation chain there must occur a `reference-frame type' interaction of this general form. If the direct \(P\)-\(S\) interaction is not of this type, then a preceding preparer is supposed to provide the relevant reference-frame interaction.

\subsection{Chain of preparers}
\label{GP}
Since the state of the preparer itself must be a coherent superposition, the authors further address the question of the ``grand-preparer'' $GP$, namely the system that prepared the preparer itself. They show that, once the preparer associated with the measured system has been identified, the grand-preparer does not participate in the subsequent conservation balance during the measurement of the system. 

They argue as follows. Since the preparer $P$ is itself a coherent superposition of the conserved quantity, there must exist a preceding system---a grand-preparer $GP$---responsible for preparing that superposition. Incorporating this entire preparation history, the joint state of the grand-preparer, preparer, and system prior to measurement is given by:
\begin{equation}
\sum_{k,l,m=-\infty}^{\infty} \Phi_g(k) \Phi_p(l) \alpha(m) \ket{k-l}_G \ket{l-m}_P \ket{m}_S 
\end{equation}
Now, assume that the conserved quantity of the system $S$ is measured, yielding the specific outcome $l_0$. The state updates to
\begin{equation}
 \sum_{k,l=-\infty}^{\infty} \Phi_g(k) \Phi_p(l) \ket{k-l}_G \ket{l-l_0}_P\ket{l_0}_S 
\end{equation}
By calculating the marginal probability distributions from this post-measurement state, the authors show that the distribution of the conserved quantity for the preparer changes by exactly $-l_0$, offsetting the change observed in the system. Crucially, however, the distribution for the grand-preparer remains unchanged. The grand-preparer exchanged angular momentum with the preparer to generate its initial state, but the grand-preparer’s marginal distribution remains unchanged. It is argued that this is a conceptually vital feature, as otherwise, if the shift propagated endlessly backward through the historical chain of grand-preparers, verifying conservation would require measuring the state of the entire universe, which would render the conservation law physically meaningless and unfalsifiable.
 
\subsection{Extension to several noncommuting conserved quantities}
\label{SN}
 
The authors acknowledge that their analysis is restricted to the simplest case: the angular momentum of a particle moving on a circle, where measuring the relevant component does not disturb other conserved quantities. They nevertheless conjecture that their proposal can be extended to situations involving several noncommuting conserved quantities.

For three-dimensional angular momentum, for example, they suggest that when $L_z$ is measured, the preparer supplies the compensating $z$-component required for conservation in each individual outcome. At the same time, because the measurement disturbs $L_x$ and $L_y$, the measuring device exchanges those components with the system so that they too remain conserved. The authors do not provide an analysis of this more general case, leaving it instead for future work. 
 
\section{Assessing the proposed framework}
\label{As}

In this section, we examine the framework proposed by \cite{collins2025conservation}, focusing first on their responses to the four questions introduced at the end of Section~\ref{prop}. We argue that these responses raise significant conceptual difficulties. We then argue that, independently of these justificatory issues, the proposed mechanism itself fails to establish energy conservation.

\subsection{Reference frames to the rescue?}

After describing the proposal, \cite{collins2025conservation} asks what happens if a system is simply ``given'' in a pure, isolated superposition of eigenstates of a conserved quantity, with no residual entanglement with any preparer. The response relies crucially on the role of reference frames. Their claim is that a superposition of angular-momentum eigenstates cannot be specified absolutely, since the relative phases between its components encode angular information that is meaningful only relative to some frame. From this they infer that a physical reference system must always be involved, and ultimately identify that system with the preparer.

However, such an argument appears to rely on a substantive relationalist thesis about the physical implementation of reference frames, while presenting it as if it followed directly from the claim that absolute angle is meaningless. The latter claim is relatively uncontroversial: descriptions related by a global rotation need not correspond to distinct physical situations, and angular quantities are defined only relative to a choice of frame. But this is a statement about the representational structure of the theory. It does not by itself imply that the reference frame must be represented by an additional material system that forms part of the physical state of the world and dynamically interacts with the system under consideration. To move from the absence of an absolute angle to the necessity of a physical reference system is therefore to introduce an additional ontological and dynamical assumption.

The same point can be seen in an ordinary Galilean setting. The position of a particle is frame-dependent, and there is no privileged spatial origin relative to which its coordinates must be specified. Yet we would not normally conclude that a universe containing a single particle is physically ill-defined because there is no second material system relative to which its position can be measured. Rather, one may describe the particle using an arbitrary coordinate system, with descriptions related by spatial translations understood as representing the same physical situation. The fact that position is not absolute therefore does not force us to introduce another physical object as a reference body. By the same token, the fact that absolute angle is not physically meaningful does not by itself establish that every superposition of angular-momentum states must be defined relative to a further quantum system playing the role of a material reference frame.

Moreover, suppose one nevertheless accepted the inference that, given a particle in Galilean spacetime, a second physical system must be introduced in order for its position to be meaningful, or, analogously, that a particle moving on a circle in a superposition of angular-momentum eigenstates requires an additional physical system to provide an angular reference frame. The same reasoning could then be applied to the enlarged system consisting of the original particle and its reference system. In the Galilean case, the center of mass of this composite system itself has a position, which, according to the same argument, would require a further physical system relative to which that position is defined. Likewise, in the angular-momentum case, if the composite system is itself in a state whose angular properties require a reference frame, yet another physical system would have to be introduced. Repeating the argument generates an infinite regress of reference systems. Applied to the classical example, it would lead to the striking conclusion that Newtonian mechanics cannot consistently describe a universe containing any finite number of particles, since an additional particle would always be required to supply the reference frame for the system already under consideration. This strongly suggests that the original inference is mistaken: the need to specify a reference frame in describing a system does not, by itself, entail the existence of an additional material system physically realizing that frame.

We therefore conclude that the response offered by \cite{collins2025conservation} to the question of why a system cannot simply be given in a pure superposition is not satisfactory. The fact that certain quantities are meaningful only relative to a reference frame does not by itself imply that such a frame must be physically instantiated by an additional system. Without a further argument establishing this stronger claim, the proposed resolution of point (i) remains unsupported.

\subsection{Is the preparation interaction well justified?}

We now turn to the second question: what guarantees that the particular preparation mechanism required by the proposal actually occurs? As we saw in the previous subsection, the authors of \cite{collins2025conservation} appeal to the role of reference frames in answering this question. There we argued that the inference from the need for a reference frame to the existence of an additional physical reference system is itself problematic. Here, however, we set that objection aside. Even granting the authors' reference-frame assumptions, a further difficulty remains: one must distinguish between showing that the proposed interaction can realize the desired preparation and showing that some interaction of this type must occur in the preparation history of every relevant superposition.

The authors consider an interaction of the form
\begin{equation}
|\Phi\rangle_P|0\rangle_S
\longrightarrow
\sum_m \Psi(m)\,|\Phi-m\rangle_P|m\rangle_S ,
\end{equation}
which correlates every component in which the system acquires an amount $m$ of the conserved quantity with a preparer state shifted by the opposite amount. By construction, this interaction preserves the total conserved quantity. Moreover, as discussed above, in the angular-momentum case the authors show that the same interaction can be interpreted as preparing the desired state relative to the preparer, which thereby plays the role of a physical reference frame. Thus, they show that this interaction provides a possible conservation-preserving implementation of the desired relational state.

For the general conservation argument, however, possibility is not enough. What is required is the much stronger claim that, for any physically realizable superposition of eigenstates of a conserved quantity, there must have occurred somewhere in its preparation history a reference-frame-type interaction that generates the compensating correlations on which the proposal relies. Otherwise, the same superposition might be produced by a different conservation-preserving interaction without leaving behind the required residual correlations with a preparer. Indeed, the authors explicitly acknowledge that their interaction is not the most general conservation-preserving one and consider cases in which the desired superposition is transferred to the system without leaving the relevant entanglement with the immediate preparer.

The crucial gap is therefore between viability and necessity. Showing that a particular reference-frame-type interaction can prepare the desired state while preserving the conserved quantity does not establish that every physically admissible preparation must contain such an interaction at some stage. Yet this is precisely what the general proposal requires. At the decisive point, the authors of \cite{collins2025conservation} do not prove this necessity; they explicitly \emph{postulate} that a reference-frame-type interaction must occur somewhere in the preparation chain, even if not in the immediate interaction between the system and its preparer. Thus, even granting the reference-frame considerations of the previous subsection, the proposal still relies on an additional assumption: that every relevant preparation history must instantiate the specific kind of compensating structure required by the conservation argument.

\subsection{Chain of preparers, conservation localization, and statistical versus non-statistical conservation}

We start this subsection by noting that, in \cite{collins2025conservation}, it is argued that, once the role of the preparer is properly taken into account, the proposed mechanism recovers a stronger, non-statistical form of quantum conservation, rather than the merely statistical conservation usually associated with quantum mechanics. There is, however, a difficulty with this claim. For the mechanism to reproduce the usual description of the prepared system as approximately pure, the initial state of the preparer cannot be an eigenstate of the conserved quantity. If it were, the states of the preparer correlated with the different components of the system would be orthogonal rather than nearly indistinguishable, and the reduced state of the system would fail to approximate the intended pure superposition. The preparer must therefore itself begin in a superposition of different values of the conserved quantity.

Consequently, according to the proposal itself, the conserved quantity does not possess a well-defined value in the initial state of the preparer. Nor does this change after a particular measurement outcome is obtained: the corresponding final state of the preparer is merely a shifted version of its initial superposition and is therefore, once again, not an eigenstate of the conserved quantity. Thus, neither before nor after the measurement does the preparer possess a definite value that can straightforwardly be said to decrease by precisely the amount gained by the measured system. In the sense emphasized in \cite{maudlin2020status}, this falls short of conservation in the usual classical sense, according to which some definite physical quantity possesses a value that remains unchanged, or is transferred between systems while the total definite value is preserved.

What the mechanism is argued to establish is a correspondence between the relevant probability distributions: the shift in the preparer's distribution compensates for that associated with the measured system, so that the appropriate statistical properties of the total conserved quantity are preserved. But this is precisely the kind of statistical conservation already familiar from quantum mechanics. The proposal may therefore succeed in extending standard quantum statistical conservation to individual measurement scenarios, in the sense of retaining the compensating correlations when one conditions on a particular outcome. It does not follow, however, that it thereby recovers a genuinely stronger, non-statistical notion of conservation. The quantity assigned to the preparer remains indefinite throughout, so the claim that a definite amount has been transferred from one subsystem to the other is not established by the formalism itself.

This observation also puts pressure on the claim that the conservation mechanism can be fully localized to the system--preparer pair. The authors' point is that, once the preparer is included, no appeal to additional systems is required to account for the compensating change associated with a particular measurement outcome. However, if the mechanism establishes only statistical conservation, the sense in which this accounting is localized is correspondingly limited. In a given measurement outcome, the system acquires a definite value of the conserved quantity, but the preparer remains in a superposition and therefore possesses no definite value that can be said to have changed by the compensating amount. What closes within the system--preparer pair is instead the relevant probability distribution: the distribution associated with the preparer is shifted so as to compensate for the outcome obtained by the system. Thus, the proposal does seem to show that the statistical bookkeeping can be confined to the system and preparer, but it is less clear that it establishes the localized, outcome-by-outcome conservation that motivates the proposal.

Returning to the state of the preparer, consider the full prepared state
\begin{equation}
\ket{\Psi'}=\sum_n \alpha(n)\ket{\Phi-n}_P\otimes\ket{n}_S.
\end{equation}
For the reduced state of the system to approximate the target pure state, the shifted preparer states corresponding to components with non-negligible amplitude must be nearly indistinguishable:
\begin{equation}
\langle\Phi-n|\Phi-m\rangle_P\simeq 1.
\end{equation}
This condition can be satisfied when the preparer has a sufficiently broad distribution of the conserved quantity. For example, if the conserved quantity is linear momentum, the preparer must have a large momentum uncertainty.

For concreteness, suppose the preparer is described by a Gaussian wave packet with uncertainty $\Delta_P$ in the variable conjugate to the conserved quantity. In the basis of the conserved quantity, its distribution then has a width $\sigma_P$ inversely proportional to $\Delta_P$. For a suitable choice of Gaussian convention, the overlap between shifted preparer states takes the form
\begin{equation}
\langle\Phi_P-n|\Phi_P-m\rangle
\sim
\exp\left[-\frac{(n-m)^2}{8\sigma_P^2}\right].
\end{equation}
Defining
\begin{equation}
\sigma_S:=\max |n-m|
\end{equation}
over the significantly populated components of the superposition, the condition
$\langle\Phi_P-n|\Phi_P-m\rangle\simeq1$ requires
\begin{equation}
\sigma_P\gg\sigma_S.
\end{equation}

Now, given that the state of $P$ itself is a superposition of eigenstates of a conserved quantity, it must also be subjected to the same reasoning. And, since the distribution of $P$ in the conserved quantity has width $\sigma_P$, preparing $P$ in an approximately pure state by the same mechanism requires a grand-preparer whose corresponding width satisfies
\begin{equation}
\sigma_{GP}\gg\sigma_P.
\end{equation}
Applying the argument recursively generates a hierarchy
\begin{equation}
\sigma_S\ll\sigma_P\ll\sigma_{GP}\ll\sigma_{GGP}\ll\cdots,
\end{equation}
and therefore, for the corresponding conjugate uncertainties,
\begin{equation}
\Delta_P\gg\Delta_{GP}\gg\Delta_{GGP}\gg\cdots.
\end{equation}

Thus, by applying the approximate-purity requirement recursively at each stage of the preparation chain, successive preparers must have increasingly broad distributions of the conserved quantity. In the case of linear momentum, for example, every successive preparer must possess a broader momentum distribution. The recursive application of the proposal therefore appears to require an increasingly idealized hierarchy of reference systems, eventually involving extreme physical scales and, depending on the conserved quantity under consideration, potentially severe resource or energetic requirements.

A final argument concerning a chain of preparers, in the spirit of the discussion above concerning the strange implication that Newtonian physics would require an infinite number of particles, is the following. According to \cite{collins2025conservation}, a system $S$ that appears to be in a pure superposition of different values of a conserved quantity, such as Eq.~(\ref{Sup}), cannot simply be in such a state: without a physical reference system, the superposition is claimed not even to be physically meaningful. It must therefore in reality be described as part of an entangled state of $S$ and a preparer $P$, such as the state on the right-hand side of Eq.~(\ref{UT}). But this enlarged system $P+S$ is itself described by a pure state and, since the preparer must itself be in a superposition of the conserved quantity, the composite system will in general be in a superposition of different values of the corresponding total conserved quantity. The authors' own principle therefore applies again: this pure superposition cannot simply be taken as the state of $P+S$, but must instead be understood relative to a further physical system, a grand-preparer $GP$, with which the composite system is entangled. The same conclusion then follows for the enlarged system $GP+P+S$, and so on. There appears to be no point at which this reasoning can terminate. Taken at face value, the claim that such pure superpositions are meaningless without an additional physical reference system therefore entails that any apparently pure superposition of this kind is, in reality, part of an entangled state involving an infinite hierarchy of further physical systems.\footnote{Notice that this regress is not blocked by the authors’ observation that the grand-preparer does not enter the conservation balance for a measurement on \(S\): the present argument concerns the physical meaningfulness of the enlarged pure state itself, rather than the propagation of the compensating shift.}

\subsection{On the extension to several conserved quantities}

Can the proposal be extended to several noncommuting conserved quantities?\footnote{Rotational symmetry itself makes the problem of noncommuting conserved quantities unavoidable. If angular momentum is conserved as a consequence of rotational invariance, then all three components, $J_x$, $J_y$, and $J_z$, must be treated on an equal footing. Since these observables do not commute with one another, the difficulty discussed here arises already in the most elementary setting of angular-momentum conservation.} We identify significant challenges in extending the framework beyond the case of a single conserved observable. Although the authors explicitly restrict their analysis to this simplified setting and merely conjecture that their proposal can be generalized to situations involving several noncommuting conserved quantities, leaving the issue for future work, we argue that such an extension appears highly problematic.

Consider a scenario involving several noncommuting conserved quantities, $\hat{A}$, $\hat{B}$, etc., and a system $S$ prepared in a state $|\psi\rangle_S$. In general, the same state can be expressed as a superposition in the eigenbases of the different conserved observables:
\begin{equation}
|\psi\rangle_S
=
\sum_n \alpha_n \ket{a_n}_S
=
\sum_n \beta_n \ket{b_n}_S
=
\cdots .
\end{equation}

According to the proposed framework, a superposition of eigenstates of a conserved quantity should in fact be understood as part of an enlarged state of the form appearing on the right-hand side of Eq.~(\ref{UT}), in which the different components of the system are correlated with compensating states of a preparer. The problem is that a preparation interaction constructed to produce the required correlations for one conserved observable, say $\hat{A}$, will not in general produce the corresponding compensating correlations for another conserved observable $\hat{B}$ that does not commute with $\hat{A}$.

For instance, consider a spin-$1/2$ system that is to be prepared in the state
\begin{equation}
\ket{\Psi}_S
=
\frac{1}{\sqrt{2}}
\left(
\ket{+z}_S+\ket{-z}_S
\right).
\end{equation}
Following the logic of \cite{collins2025conservation}, suppose the preparer is another spin-$1/2$ system and that, prior to preparation, the joint state is
\begin{equation}
\ket{\Psi_0}
=
\ket{+z}_P\ket{-z}_S .
\end{equation}
A preparation interaction conserving total $z$-spin can then produce
\begin{equation}
\ket{\Psi'}
=
\frac{1}{\sqrt{2}}
\left(
\ket{-z}_P\ket{+z}_S
+
\ket{+z}_P\ket{-z}_S
\right).
\end{equation}
Each branch has the same total $z$-spin as the initial state, so the preparation has precisely the compensating structure required by the proposal for $S_z$.

Consider, however, the same process from the perspective of $S_x$. Before the preparation interaction,
\begin{equation}
\ket{\Psi_0}
=
\frac{1}{2}
\left(
\ket{+x}_P\ket{+x}_S
-
\ket{+x}_P\ket{-x}_S
+
\ket{-x}_P\ket{+x}_S
-
\ket{-x}_P\ket{-x}_S
\right),
\end{equation}
whereas after preparation,
\begin{equation}
\ket{\Psi'}
=
\frac{1}{\sqrt{2}}
\left(
\ket{+x}_P\ket{+x}_S
-
\ket{-x}_P\ket{-x}_S
\right).
\end{equation}
The Born-rule distribution of the total $x$-spin is therefore not preserved by the preparation interaction. The mechanism that establishes the required compensation for $S_z$ does not simultaneously establish it for $S_x$.

Thus, the framework appears capable, at most, of guaranteeing conservation for the particular observable singled out by the preparation interaction. More fundamentally, however, this raises a question about how the proposal itself is to be applied. A state such as $|\psi\rangle_S$ is simultaneously a superposition in the eigenbases of several noncommuting conserved observables. If the proposal tells us that such a state must really be understood as an enlarged system--preparer state containing compensating correlations, which conserved observable determines the relevant correlations? The structure required for one observable will not, in general, be the structure required for another, and the framework provides no general prescription for selecting among them or for realizing all of them simultaneously.

At this point, the authors seem to suggest that the interaction with the measuring device could come to the rescue. We find this move somewhat surprising, since the central conceptual lesson motivating \cite{collins2025conservation} is precisely that, in order to understand conservation in the quantum domain, one should not look primarily to the measuring device but to the system responsible for preparing the initial state. The proposed role of the preparer was introduced precisely because the measuring apparatus was not supposed to account for the apparent change in the conserved quantity associated with an individual outcome. Appealing once again to the measuring device when additional conserved quantities are considered therefore seems to weaken the very distinction on which the proposal was originally based.

In sum, the extension to several noncommuting conserved quantities faces two related difficulties. First, a preparation interaction that produces the compensating correlations required to conserve one observable will not, in general, do so for incompatible observables. Second, since a given quantum state can be expressed as a superposition in the eigenbases of several such observables, the framework provides no general prescription for determining which corresponding system--preparer correlation structure is physically realized. The appeal to the measuring device does not resolve this tension and appears to weaken the central motivation for introducing the preparer in the first place.

\subsection{Additivity and energy conservation}

To conclude this section, we show that, independently of the problems discussed above, the mechanism proposed in \cite{collins2025conservation} fails to establish energy conservation.

As we saw, the proposal holds that any superposition of eigenstates of a conserved quantity must ultimately be prepared through interaction with another physical system, which remains correlated with the system of interest. Once this preparer is included in the description, the change associated with a particular measurement outcome is claimed to be balanced by an opposite change in the preparer, thereby yielding conservation in each individual outcome.

The proposed mechanism relies crucially on the additivity of the conserved quantity. The compensating structure is supposed to work because a change in the quantity associated with the measured system can be exactly balanced by an opposite change in the preparer, so that the total quantity is obtained simply by summing the contributions associated with the two subsystems. Without such an additive decomposition, the shifts assigned to the system and preparer would not by themselves suffice to establish conservation.

This assumption becomes nontrivial, however, once the preparation process is taken seriously as a physical interaction. The system and preparer must interact in order to transform an initially separable state into the entangled state on which the proposed explanation depends. For some conserved quantities, such as angular momentum, an additive total quantity can remain well defined even in the presence of interactions. Energy is different in an important respect: during the preparation interaction the total Hamiltonian generally takes the form
\begin{equation}
H_{\mathrm{tot}}=H_S+H_P+H_{\mathrm{int}},
\end{equation}
so that the total energy cannot in general be identified with the sum of the energies assigned separately to the system and preparer. In such cases, demonstrating opposite shifts in the subsystem energies is therefore insufficient to establish conservation; the contribution of the interaction itself must also be included. This raises a difficulty for extending the proposed bookkeeping mechanism straightforwardly to energy conservation.

\section{Conclusion}
\label{Con}

We have examined the proposal of \cite{collins2025conservation} according to which conservation laws can be restored at the level of individual quantum measurement outcomes by including the physical system responsible for preparing the measured state. The proposal highlights an important and often neglected point: the preparation history of a quantum state may matter when questions of conservation are considered. Nevertheless, we have argued that the stronger conclusions drawn from this observation are not established.

At the level of justification, the framework relies on substantive assumptions that do not follow from the reference-frame considerations invoked in their support. In particular, the necessity of a physical reference system and of a reference-frame-type preparation interaction is assumed rather than demonstrated, while recursive application of the same reasoning generates further difficulties and the extension to several noncommuting conserved quantities remains problematic. Moreover, even granting the framework, the resulting mechanism appears to preserve only statistical properties of the relevant conserved quantity rather than establishing conservation of a definite quantity in the stronger, non-statistical sense claimed. The claim of locality is correspondingly weakened, and in the case of energy the interaction term required for the preparation process prevents the total energy from being identified simply with the sum of the subsystem energies. We therefore conclude that the proposal does not establish conservation in every quantum measurement outcome in the general sense claimed.

\bibliographystyle{ieeetr}
\bibliography{ref.bib}
\end{document}